\documentclass[twocolumn,floats,showpacs,amsmath,amssymb,pre]{revtex4}
\usepackage{graphicx}
\begin{document}

\title{Antiferromagnetic Ising model in scale-free networks}
\author{Carlos P. Herrero}
\affiliation{Instituto de Ciencia de Materiales de Madrid,
         Consejo Superior de Investigaciones Cient\'{\i}ficas (CSIC),
         Campus de Cantoblanco, 28049 Madrid, Spain }
\date{\today}

\begin{abstract}
The antiferromagnetic Ising model in uncorrelated scale-free networks 
has been studied by means of Monte Carlo simulations. 
These networks are characterized by a connectivity (or degree)
distribution $P(k) \sim k^{-\gamma}$.
The disorder present in these complex networks frustrates the
antiferromagnetic spin ordering,
giving rise to a spin-glass (SG) phase at low temperature.
The paramagnetic-SG transition temperature $T_c$
has been studied as a function of the parameter $\gamma$ and the
minimum degree present in the networks.
$T_c$ is found to increase when the exponent $\gamma$ is reduced,
in line with a larger frustration caused by the presence of nodes with 
higher degree.
\end{abstract}
\pacs{64.60.De, 05.50.+q, 75.10.Nr, 89.75.Hc}
%
% 64.60.De Statistical mechanics of model systems (Ising model, Potts model,
%          field-theory models, Monte Carlo techniques, etc)
% 05.50.+q Lattice theory and statistics (Ising, Potts, etc.)
% 75.10.Nr Spin-glass and other random models
% 89.75.Hc Networks and genealogical trees

\maketitle

\section{Introduction}

 Several types of natural and artificial systems have a network structure,
where nodes represent typical system units and edges play the role of 
interactions between connected pairs of units. 
This kind of description of complex systems as networks or graphs has
attracted much interest in recent years.
Thus, complex networks have been used to model various types of real-life
systems (biological, social, economic, technological), and to study several
processes taking place on them \cite{al02,ne03,ne06,do03a,co07}.
Some models of networks have been designed to explain
empirical data in various fields. This is the case of the
so-called small-world \cite{wa98} and scale-free
networks \cite{ba99b}, which incorporate different aspects of real systems.

In scale-free (SF) networks the degree distribution $P(k)$, where $k$ is the
number of links connected to a node, has a power-law decay $P(k) \sim
k^{-\gamma}$ \cite{do02a,go02}.
This kind of networks have been found in the internet \cite{si03},
in the world-wide web \cite{al99}, for protein interactions \cite{je01},
and in social systems \cite{ne01a}.
The origin of such power-law degree distributions was addressed by
Barab\'asi and Albert \cite{ba99b}, who found that two ingredients are
sufficient to explain the scale-free character of networks:
growth and preferential attachment. They concluded that the combination
of these criteria yields non-equilibrium SF networks with an exponent 
$\gamma = 3$.
One can also consider equilibrium SF networks, defined as statistical 
ensembles of random networks with a given degree distribution
$P(k)$ \cite{do02a,bo06}, for which it is possible to analyze 
various properties as a function of the exponent $\gamma$.
 
Cooperative phenomena in complex networks display unusual 
characteristics due to their peculiar topology 
\cite{ba00,sv02,he02a,lo04,ca06,do08}. 
In particular, the ferromagnetic (FM) Ising model has been thoroughly 
studied in scale-free networks \cite{ig02,do02b,le02,he04},
and its critical behavior was found to be dependent on the exponent 
$\gamma$. For finite $\langle k^2 \rangle$, a ferromagnetic 
to paramagnetic transition occurs at a finite temperature $T_c$.
However, when $\langle k^2 \rangle$ diverges 
(as happens for $\gamma \leq 3$), the spin system remains in an ordered 
FM phase at any temperature, so that no phase transition appears in the
thermodynamic limit.

Here we study the antiferromagnetic (AFM) Ising model in equilibrium 
(uncorrelated) scale-free networks with several values of the exponent 
$\gamma$. This model contains the two basic ingredients necessary to 
produce a spin-glass (SG) phase at low temperature: disorder and 
frustration. 
In some spin-glass models, such as the Sherrington-Kirkpatrick model, 
all spins are assumed to be mutually connected \cite{my93,fi91}, whereas
in others random graphs with finite (low) connectivity are considered
\cite{ka87,de01,ki05,bo03}.
 For the AFM Ising model on scale-free networks, we expect 
to find features intermediate between these two cases.

Spin glasses on complex networks have been studied in recent years 
by using several techniques, such as transfer matrix 
analysis \cite{ni04}, replica symmetry breaking \cite{we05},
defect-wall calculations \cite{we07}, and an effective field theory
\cite{os08}.
In this paper, we employ Monte Carlo (MC) simulations to study the
paramagnetic to spin-glass phase transition appearing in scale-free
networks. In this line, MC simulations have been carried out earlier
to analyze spin-glass phases appearing for the AFM Ising model
in Barab\'asi-Albert scale-free networks \cite{ba06}, as well as in
small-world networks \cite{he08}.    

The paper is organized as follows.
In Sec.~II we describe the networks and the computational method
used here. 
In Sec.\,III we present results for the heat capacity, energy, and 
spin correlation, as derived from MC simulations. 
In Sec.\,IV we characterize the spin-glass phase through the overlap 
parameter and transition temperature. 
The paper closes with the conclusions in Sec.\,V.

\section{Model and method}

We consider SF networks with degree distribution $P(k) \sim k^{-\gamma}$.
They are characterized, apart from the exponent $\gamma$ and the system 
size $N$, by the minimum degree $k_0$. 
We assume that $P(k) = 0$ for $k < k_0$.
Our networks are uncorrelated, in the sense that degrees of
nearest neighbors are statistically independent.
This means that the distribution $P(k,k')$ of degrees of nearest-neighbor
nodes fulfills the relation \cite{do03a}
\begin{equation}
  P(k,k') = \frac{k \, k'}{\langle k \rangle^2}  P(k) P(k') \, .
\label{kk1}
\end{equation}
Alternatively, one can use a correlation coefficient $r$ defined as
\begin{equation}
 r = \frac{ \langle k \, k'\rangle - \langle k \rangle \langle k' \rangle }
           {\sigma_k^2} \, ,       
\label{rr}
\end{equation}
where the averages are taken over all links and $\sigma_k^2$ is the
variance of the degree distribution. This coefficient $r$ is zero for
uncorrelated networks.

For the numerical simulations we have generated networks with several
values of $\gamma$, $k_0$, and $N$.
To generate a network, we first define the number of nodes $N_k$ with 
degree $k$, according to the distribution $P(k)$, which can be 
conveniently done by using the so-called transformation method 
\cite{ne05}.  Then, we ascribe a degree to each node according to 
the set $\{N_k\}$, and finally connect at random ends of links 
(giving a total of $L = \sum_k k N_k/2$ connections), 
with the conditions:
(i) no two nodes can have more than one bond connecting them, and
(ii) no node can be connected by a link to itself.
We have checked that networks generated in this way are uncorrelated, i.e.
they fulfill Eq.~(\ref{kk1}), and $r = 0$.
The networks considered here contain a single component, i.e. any node in a
network can be reached from any other node by traveling through a finite
number of links.

Given a network with a particular set of links,
we consider an Ising model with the Hamiltonian:
\begin{equation}
H = \sum_{i < j} J_{ij} S_i S_j   \, ,
\end{equation}
where $S_i = \pm 1$ ($i = 1, ..., N$), and the coupling matrix
$J_{ij}$ is given by
\begin{equation}
J_{ij}  \equiv \left\{
     \begin{array}{ll}
         J (> 0), & \mbox{if $i$ and $j$ are connected,} \\
         0, & \mbox{otherwise.}
     \end{array}
\right.
\label{Jij}
\end{equation}
This means that each edge in the network represents an AFM interaction
between spins on the two linked nodes.
Note that, contrary to the usually studied models for spin glasses
in which both FM and AFM couplings are present,
in our model all couplings are antiferromagnetic
(similarly to Refs. \cite{he08,kr05}).

For a given network, we carried out Monte Carlo simulations at several
temperatures, sampling the spin configuration space by the Metropolis update 
algorithm \cite{bi97}, and using a simulated annealing procedure. 
Several variables characterizing the spin system
have been calculated and averaged for different values of the
considered parameters.
For each set of parameters ($\gamma$, $k_0$, $N$), 1000 network realizations
were generated, and the largest networks included 16000 nodes. 
In the sequel, we will use the notation $\langle ... \rangle$ to
indicate a thermal average for a network, and $[ ... ]$ for an average
over networks with a given parameter set.

\section{Thermodynamic variables}

We present first results for the heat capacity per site, $c_v$, 
as a function of temperature for several values of the exponent
$\gamma$. $c_v$ has been derived from the energy fluctuations 
$\Delta E$ at a given temperature, by using the expression
\begin{equation} 
c_v = \frac {[ (\Delta E)^2 ]} {N T^2}  \,  ,
\end{equation} 
where $(\Delta E)^2 = \langle E^2 \rangle - \langle E \rangle^2$.
We have checked that the results obtained in this way coincide within 
statistical noise with those derived by calculating the heat capacity as 
the energy derivative $[d \langle E \rangle / d T] / N$.
Note that we take the Boltzmann constant $k_B = 1$.

\begin{figure}
\vspace{-2.0cm}
\includegraphics[width= 9cm]{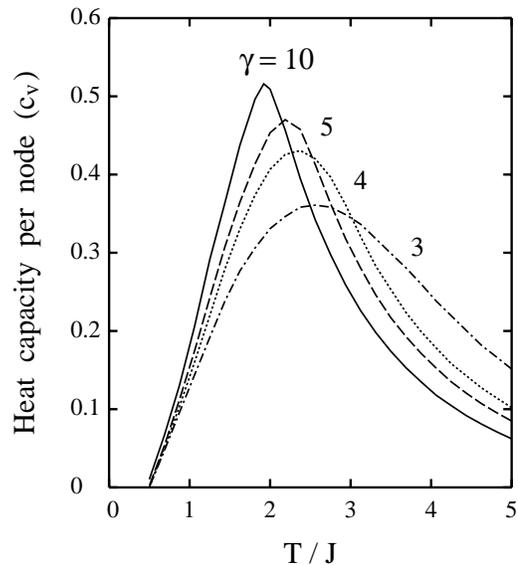}
\vspace{-2.5cm}
 \caption{
Heat capacity per node, $c_v$, vs temperature for scale-free
networks with $N$ = 8000 nodes and minimum degree $k_0 = 5$.
The plotted curves correspond to different values of the exponent
$\gamma$: 10 (solid line), 5 (dashed line), 4 (dotted line), and
3 (dashed-dotted line).
} \label{fig1} \end{figure}

The temperature dependence of $c_v$ is displayed in Fig.~\ref{fig1} for
scale-free networks with various values of $\gamma$ between 3 and 10. 
The data shown correspond to networks including 8000 nodes.
For increasing $\gamma$, we observe that the maximum of $c_v$ shifts to 
lower $T$, and the peak becomes narrower. 
This narrowing is in line with that observed earlier for the heat capacity
in the AFM Ising model on small-world networks, when the disorder
is reduced \cite{he08}.
In our case of scale-free networks, larger values of the exponent
$\gamma$ correspond to networks with a higher homogeneity in the node
connectivities (less dispersion in the degree distribution), causing
a narrower peak in the heat capacity, as shown in Fig.~1.
The peak shift to lower temperatures suggests a transition from a 
paramagnetic to a SG phase at a temperature $T_c$ that decreases as 
the exponent $\gamma$ rises.
For increasing $\gamma$, one reduces the presence of nodes with a large
degree, which in turn reduces the degree of frustration in the spin
distribution (see below).

\begin{figure}
\vspace{-2.0cm}
\includegraphics[width= 9cm]{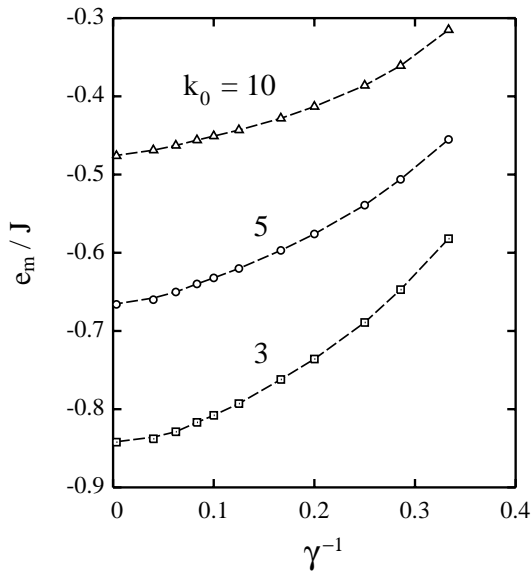}
\vspace{-2.5cm}
\caption{
Minimum energy per link derived from our simulations for the AFM Ising
model on scale-free networks, plotted as a function of the inverse
exponent $\gamma^{-1}$.
Symbols represent energy values obtained in the large-network limit
$N \to \infty$
for several values of the exponent $\gamma$, and minimum degree:
$k_0 = 10$ (triangles); $k_0 = 5$ (circles); $k_0 = 3$ (squares).
Symbols at $\gamma^{-1} = 0$ correspond to regular random networks
with a constant degree.
Lines are guides to the eye.
} \label{fig2} \end{figure}

AFM ordering on the considered random networks with power-law
distribution of degrees is frustrated by the disorder in the link 
configuration, and in particular by the presence of loops with an
odd number of nodes.
The degree of frustration can be quantified by looking at
the low-temperature energy of the system, which will be higher for 
larger frustration.  
Given the parameters $\gamma$ and $k_0$ defining the scale-free networks, 
we obtain a value for the minimum energy by extrapolating to infinite 
size the minimum energy reached in our simulations of finite-$N$ 
networks.
This extrapolation has been performed by assuming a dependence of
the energy on network size of the form:
\begin{equation}
e_m(N) = e_m + \frac{A}{N^{2/3}} \,  ,
\end{equation}
where $e_m(N)$ is the energy per link for size $N$, $e_m$ is its 
limit for $N \to \infty$, and $A$ is a fit parameter.
This kind of dependence of the low-temperature energy in spin-glass
systems was proposed by Boettcher \cite{bo03}, and has been
found to be followed by the results of our calculations using
the simulated-annealing method.
We have checked that our method to obtain a minimum energy for 
the AFM Ising model on complex networks gives similar results to
those found by using extremal optimization \cite{bo03}. 
In particular, for spin glasses on random graphs with a Poisson 
distribution of connectivities, we found results very close to those 
obtained by using this technique \cite{he08}.
In any case, the energy $e_{\text m}$ found here for each parameter 
set ($k_0, \gamma$) will be an upper limit for the lowest 
energy of the system.

\begin{figure}
\vspace{-2.0cm}
\includegraphics[width= 9cm]{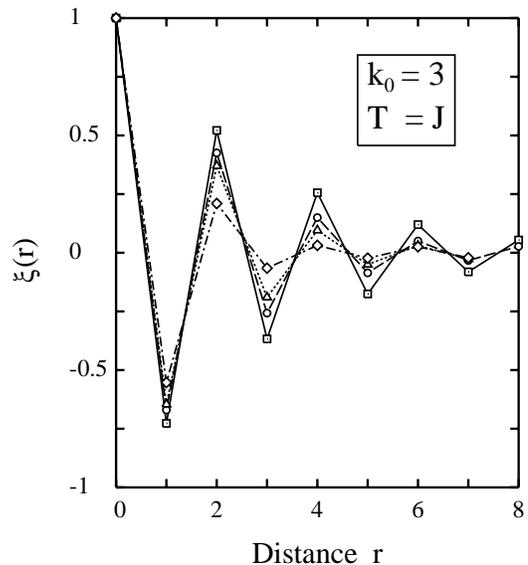}
\vspace{-2.5cm}
\caption{
Spin correlation function $\xi$ vs distance for scale-free networks
with $k_0 = 3$ and several values of the exponent $\gamma$,
at temperature $T = J$.
Symbols correspond to different values of $\gamma$:
10 (squares), 5 (circles), 4 (triangles), and 3 (diamonds).
These results were derived from simulations for SF networks including
$N$ = 8000 nodes.
} \label{fig3} \end{figure}

In Fig.~\ref{fig2} we show results for the minimum energy per link
$e_m$ found for three values of $k_0$ and several values of the exponent 
$\gamma$.  For a given $k_0$ one observes an increase in $e_m$ as
the exponent $\gamma$ is reduced. This indicates that the presence
of nodes with large degree (hubs), which is favored for small
$\gamma$, plays in our context the role of increasing the frustration
in the spin arrangement. 
For a given $\gamma$, one observes also in Fig.~\ref{fig2} an
increase in the energy $e_m$ for rising $k_0$. This shows that 
an increase in the minimum connectivity (or in the average degree
$\langle k \rangle$), causes also a higher frustration in the AFM 
ordering.

\begin{figure}
\vspace{-2.0cm}
\includegraphics[width= 9cm]{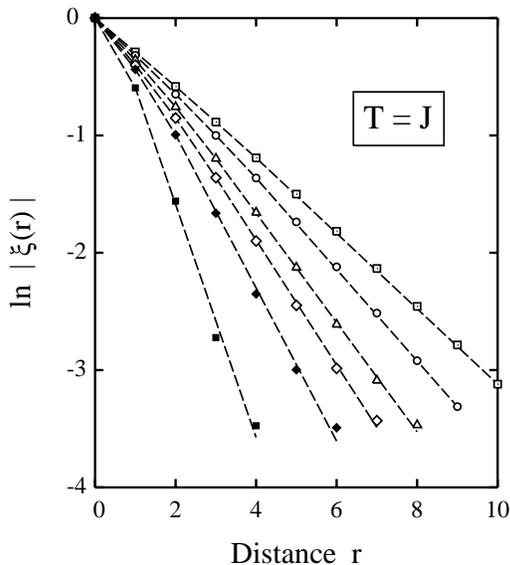}
\vspace{-2.5cm}
\caption{
Absolute value of the spin correlation function $\xi$ vs distance in
a semilogarithmic plot. Shown is $\ln |\xi(r)|$ for scale-free networks
with minimum degree $k_0 = 3$ and several values of the exponent
$\gamma$ at $T = J$.
Symbols correspond to different values of $\gamma$.
From top to bottom: $\gamma = \infty$ (regular networks, open squares),
10 (circles), 6 (triangles), 5 (open diamonds), 4 (filled diamonds), and
3 (filled squares).
Error bars are on the order of the symbol size.
These results were derived from Monte Carlo simulations for networks
including $N$ = 8000 nodes.
} \label{fig4} \end{figure}

A quantification of the short-range order present in the spin
system on scale-free networks can be obtained by calculating 
the spin correlation
\begin{equation}
 \xi(r) =  [ \langle S_i S_j \rangle_r ]   \; ,
\label{xi}
\end{equation}
where the subscript $r$ indicates that the average is taken for the
ensemble of pairs of sites at distance $r$.
Note that the dimensionless distance $r$ refers to the minimum number
of links between two nodes, also called in the literature chemical or 
topological distance.
The correlation $\xi(r)$ is shown in Fig.~\ref{fig3} for several values
of the exponent $\gamma$ at a temperature $T = J$.
This temperature is below the critical temperature $T_c$ of the
paramagnetic-SG transition for all values of $\gamma$ (see below).
As expected, $\xi(r)$ decreases faster for smaller $\gamma$,
due to the presence of nodes with large degree, and consequently
a larger frustration of the AFM ordering, as discussed above
in connection with the minimum energy per link.

To obtain more direct insight into the reduction of $\xi(r)$ with
the distance, we display in Fig.~\ref{fig4} $|\xi(r)|$ on a 
semilogarithmic plot for various $\gamma$ values.
In general, after a short transient for small $r$, 
one obtains an exponential decrease in the spin
correlation with distance, as $|\xi(r)| \sim {\rm e}^{-a r}$,
$a$ being a parameter that depends on temperature as well as on
the parameters defining the networks ($k_0$ and $\gamma$).
For the results shown in Fig.~\ref{fig4}, we find a parameter $a$
that decreases from 1.01(3) to 0.386(5) when increasing $\gamma$ from 
3 to 10, and reaches the limit $a$ = 0.321(4) for regular random networks 
with constant degree $k = k_0 = 3$.
We note that here the limit $\gamma \to \infty$ correspond to networks
(called regular \cite{bo98}), in which all nodes have the same degree
$k_0$, and consequently do not include any hub with high degree.

\section{Spin-glass behavior}

\subsection{Overlap parameter}

In the study of spin glasses, it is usual to consider two copies of
the same network, with a given realization of the disorder. 
Then, one considers a spin system on each network, both with 
different initial values of the spins, and follows their evolution
with different random numbers for generating the spin 
flips \cite{pa83,ka96}.
A particularly relevant parameter is the overlap $q$ between the two copies,
defined as
\begin{equation}
q =  \frac{1}{N}\sum_{i} S_i^{(1)} S_i^{(2)},
\label{qq}
\end{equation}
where the superscripts (1) and (2) indicate the copies. This parameter $q$
is defined in the interval $[-1,1]$, and the extreme values 1 and --1 
correspond to pairs of networks with the same spin configuration
(apart from a trivial overall flip in the --1 case).

\begin{figure}
\vspace{-2.0cm}
\includegraphics[width= 9cm]{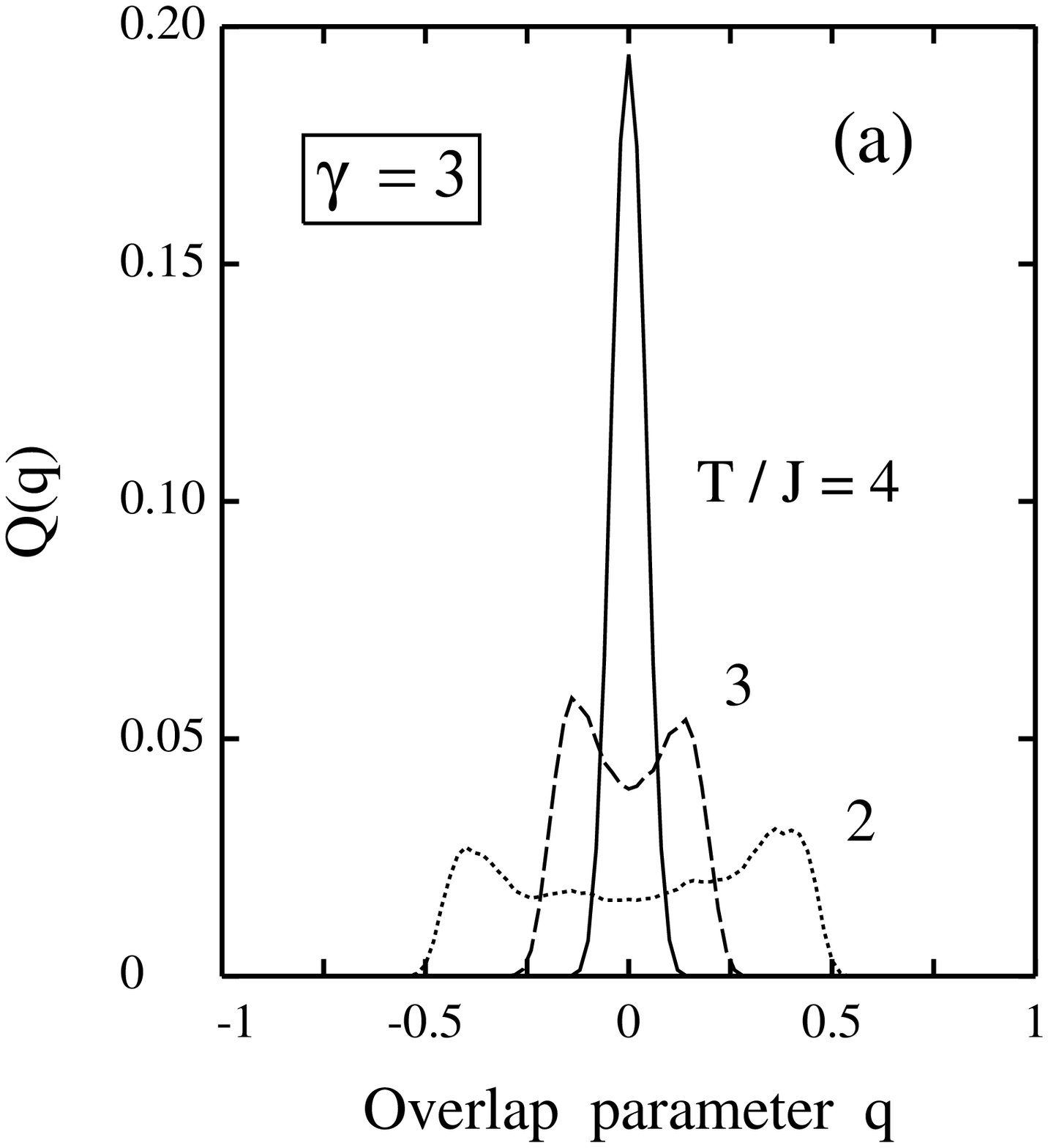}
\vspace{-2.5cm}
\end{figure}
                                                                                    
\begin{figure}
\vspace{-1.5cm}
\includegraphics[width= 9cm]{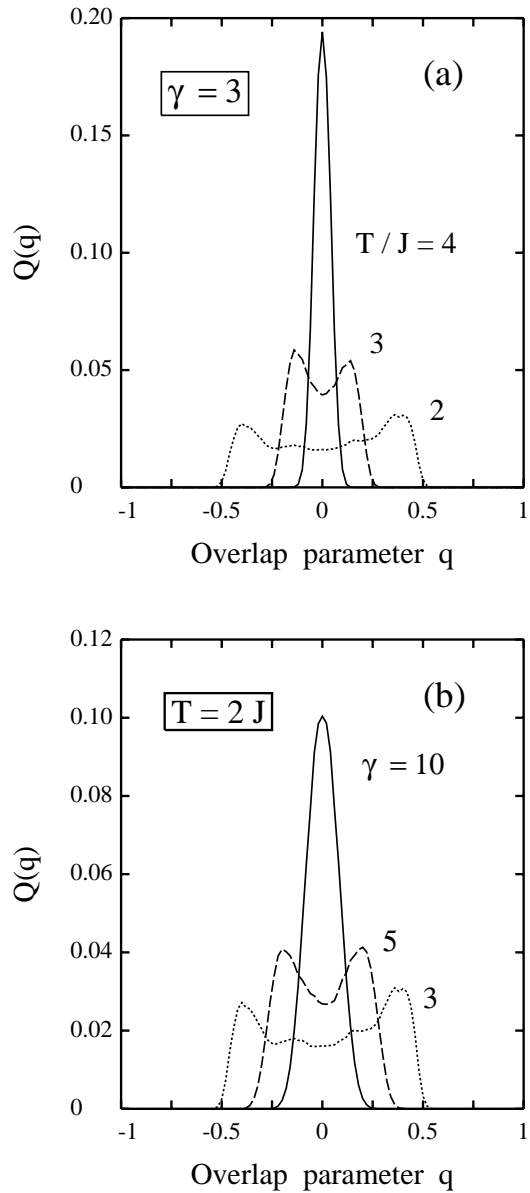}
\vspace{-2.5cm}
\caption{
Distribution of the overlap parameter $q$ for scale-free networks
with minimum degree $k_0 = 5$, as derived from MC simulations.
(a) Networks with exponent $\gamma = 3$ at three temperatures:
$T/J$ = 4, 3, and 2;
(b) networks with three different exponents $\gamma$ at temperature
$T = 2J$.
} \label{fig5} \end{figure}

We have calculated the overlap parameter $q$ for scale-free networks with
various exponents $\gamma$, and derived the probability distribution
$Q(q)$ from Monte Carlo simulations.
Results for $Q(q)$ are presented in Fig.~\ref{fig5}.
In particular, in Fig.~5(a) we display the distribution of the
overlap parameter for networks with $\gamma = 3$ at
several temperatures.
At high temperatures ($T \gg J$), the distribution $Q(q)$ shows a single 
peak centered at $q=0$, which is characteristic of a paramagnetic state. 
This peak has, however, a finite width, which results to be a 
finite-size effect. It should collapse to a Dirac delta function at 
$q=0$ in the limit $N \to \infty$.
When the temperature is lowered, $Q(q)$ broadens around $q=0$,
due to the appearance of an increasing number of frustrated links.
At still lower temperatures, frustration is more apparent, and
``freezing'' of the spins  causes the appearance of two peaks in $Q(q)$, 
symmetric respect to $q=0$, and characteristic of spin-glasses 
\cite{ka96,ka02,ba06,he08}. 
Such a distribution $Q(q)$ is associated to the break of ergodicity
occurring in the spin system at low temperatures.

In Fig.~5(b) we show the distribution $Q(q)$ for three values of the
exponent $\gamma$ at a fixed temperature $T = 2 J$. The effect of
decreasing $\gamma$ for a given $T$ is similar to that shown in 
Fig.~5(a) for lowering the temperature for a given $\gamma$, in the
sense that in both cases one passes from a high-temperature
paramagnetic phase to a spin-glass with broken ergodicity.
From the results displayed in Fig.~5(b), along with those presented in
Sect. III (specially Fig.~1 for the heat capacity and Fig.~2 for the
minimum energy per link), we expect that the freezing of the spins
in the SG phase occurs at lower $T$ for larger $\gamma$. In other words,
one expects that the transition temperature from paramagnetic to SG
will decrease for rising $\gamma$.

\subsection{Transition temperature}

The overlap parameter $q$ can be further employed to obtain precise 
values of the paramagnetic-SG transition temperature. To this end,
one can use the fourth-order Binder cumulant \cite{bi97,ka96} 
\begin{equation} 
g_N(T) = \frac{1}{2}\left(3 - \frac{ \left[\langle q^4\rangle\right]_N }
   {\left[ \langle q^2 \rangle \right]^{2}_N} \right) \, ,
\label{Binder}
\end{equation}
which is restricted to the interval [0,1].
On one side, this parameter $g_N$ vanishes at high temperatures, as 
expected for a Gaussian distribution $Q(q)$ in a paramagnetic state.
On the other side, one has $g_N = 1$ whenever the distribution $Q(q)$ 
vanishes everywhere except for $|q| = 1$,
which corresponds to the case of a single ground state, and could be
reached at low temperatures. This case ($g_N = 1$) is clearly not 
expected here due to the onset of frustration, which gives rise to the 
spin-glass phase.

\begin{figure}
\vspace{-2.0cm}
\includegraphics[width= 9cm]{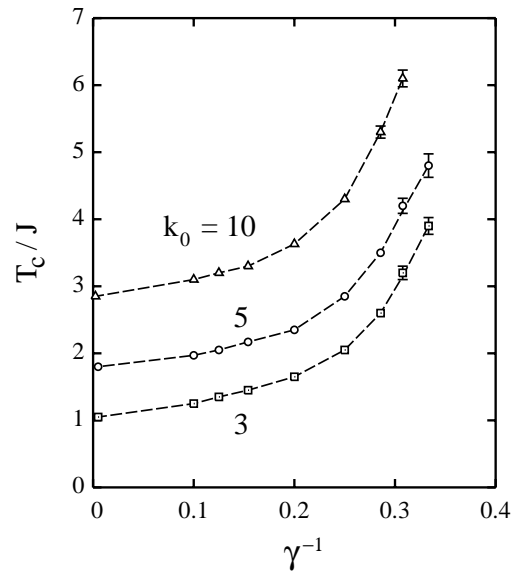}
\vspace{-2.5cm}
\caption{
Transition temperature $T_c$ from paramagnetic to SG phases, as a
function of the inverse exponent $\gamma^{-1}$, for three values
of the minimum degree $k_0$.
Error bars, when not shown, are on the order of the symbol size.
Dashed lines are guides to the eye.
} \label{fig6} \end{figure}

In general $g_N$ increases as temperature is lowered, and the
transition temperature $T_c$ can be obtained from the single crossing 
point for different network sizes $N$ \cite{bi97,he08}. 
By employing this method, we have calculated $T_c$ for scale-free
networks with several values of the parameters $k_0$ and $\gamma$, and
the results obtained are displayed in Fig.~\ref{fig6}.
In this figure, we have plotted $T_c$ as a function of $\gamma^{-1}$,
and the limit $\gamma \to \infty$ corresponds to regular random networks
with $k = k_0$. 
This limit gives a reasonable extrapolation of the $T_c$ values obtained
for $\gamma$ values up to $\gamma = 10$.
For a given minimum degree $k_0$, the transition temperature
increases as $\gamma$ is lowered.
Also, for a given value of $\gamma$, $T_c$ grows for increasing $k_0$,
similarly to the case of the minimum energy $e_m$ displayed in
Fig.~\ref{fig2}.
We note that the transition temperature $T_c$ derived from the
Binder cumulant is in the order of the maximum (negative) temperature 
derivative of the heat capacity $c_v$ for finite networks, as 
can be seen by comparing results for $k_0 = 5$ in Figs.~1 and 6.

The average degree $\langle k \rangle$ of scale-free networks can be estimated
rather accurately by replacing the sum $\sum_k k P(k)$ by an integral. 
Thus, one finds for networks with $P(k) \sim k^{-\gamma}$ and $\gamma > 2$, 
that the average degree scales as 
\begin{equation}
  \langle k \rangle  \approx  \frac {\gamma - 1}{\gamma - 2} k_0 \, .
\label{kaver}
\end{equation}
From this expression, it is clear that 
$\partial \langle k \rangle / \partial k_0 > 0$ and
$\partial \langle k \rangle / \partial \gamma < 0$.
Since rising $k_0$ or lowering $\gamma$ cause an increase in the
transition temperature, we observe that in fact a rise in $\langle k \rangle$ 
is accompanied by a rise in $T_c$.

In this context, it is interesting to compare our results for $T_c$ in
scale-free networks with those found for the paramagnetic-SG transition
in other complex networks.
In Ref.~\onlinecite{he08} it was studied the AFM Ising model on 
small-world networks generated by rewiring links in a regular lattice
\cite{wa98}. It was found that the transition temperature decreases
as the disorder (number of random connections) increases, in networks 
where the average degree was kept constant ($\langle k \rangle = 4$).
In the limit of large disorder, those networks approach random networks
with a Poisson distribution of degrees, and the transition temperature
for the AFM Ising model was found to be $T_c = 1.71 J$.
Going back to our random networks with power-law degree distribution,
we have an average degree $\langle k \rangle \approx 4$ for $k_0 = 3$
and $\gamma = 5$ [see Eq.~(\ref{kaver})]. For these networks, we find
$T_c = 1.65 J$, as shown in Fig.~6, a value close to that obtained for
small-world-type networks in the large-disorder limit and 
$\langle k \rangle = 4$. 

All this could suggest that the transition temperature $T_c$ is
proportional to the average degree $\langle k \rangle$, irrespective of
the details of the networks under consideration. This is,
however, not the case, as can be inferred directly from the results
displayed in Fig.~6. Looking for example at the data for $k_0 = 3$,
we observe that going from $\gamma = 3$ to the limit $\gamma \to \infty$,
$\langle k \rangle$ decreases by a factor of 2 (from 6 to 3). 
On the other side, $T_c$ is reduced by a factor $\approx 4$.
Since lowering $\gamma$ increases the inhomogeneity in the degree
distribution, favoring the presence of nodes with degree much larger
than $\langle k \rangle$,
we find that this inhomogeneity helps to rise the transition
temperature.

Bartolozzi {\em et al.} \cite{ba06} studied the AFM Ising model on
Barab\'asi-Albert scale-free networks, and found a paramagnetic-SG
transition temperature $T_c = 4.0(1) J$. These nonequilibrium
networks are characterized by an exponent $\gamma = 3$, and
those authors used the particular value of the minimum degree
$k_0 = 5$.
For these parameters, we find for equilibrium networks a
transition temperature $T_c = 4.8(2) J$, a value somewhat higher
than that obtained in Ref. \onlinecite{ba06}.

We note that the error bar in the transition temperature grows when
$\gamma$ is reduced. In fact, the actual value of the cumulant $g_N$ at 
the crossing point for different $N$ values (network sizes) decreases
as $\gamma$ is lowered, and is near zero for $\gamma \sim 3$.
This coincides with results shown for this cumulant in 
Ref.~\onlinecite{ba06} for Barab\'asi-Albert SF networks, where the
value of $g_N$ at the crossing point was less than 0.01. 
This means that the signal-to-noise ratio in $g_N$ becomes poor,
and one has an increasing uncertainty in $T_c$.
For networks with $\gamma < 3$, we could not find a single crossing point
for the cumulant corresponding to different network sizes, and a
transition temperature cannot be given. We note that these $\gamma$
values correspond to SF networks with diverging $\langle k^2 \rangle$.
In this respect, it is known that the ferromagnetic Ising model 
in such networks does not show a phase transition, and remains in an
ordered FM phase at any temperature in the thermodynamic limit
\cite{ig02,do02b,le02,he04}.
Something similar could happen for the AFM Ising model in these
networks. This point remains as a challenge for future research.

\section{Conclusions}

The AFM Ising model in random networks with a power-law distribution
of degrees gives rise to a spin-glass phase at low temperature.
This is a consequence of the combination of disorder in the networks
and frustration caused by the presence of loops with odd number of
links.
The overlap parameter $q$ gives us evidence of this frustration
at low temperatures. 

The transition temperature $T_c$ from the high-temperature paramagnetic
phase to the spin-glass has been studied as a function of the minimum 
degree $k_0$ and the exponent $\gamma$ in the degree distribution.
$T_c$ is found to rise for increasing $k_0$ and for lowering $\gamma$.

For a given $k_0$, both the transition temperature and the minimum energy
per link $e_m$ found from our simulations increase as the exponent
$\gamma$ is lowered. This indicates that the degree of frustration 
in the spin configurations rises with the presence of nodes with large
degree (hubs).
The same conclusion can be reached by analyzing the spin correlation
as a function of distance, which decays faster for smaller values of
$\gamma$.

\begin{acknowledgments}
This work was supported by Ministerio de Ciencia e Innovaci\'on 
(Spain) under Contract No. FIS2006-12117-C04-03.  \\
\end{acknowledgments}

%\bibliographystyle{apsrev}
%\bibliography{networks,glass}

\begin{thebibliography}{43}
\expandafter\ifx\csname natexlab\endcsname\relax\def\natexlab#1{#1}\fi
\expandafter\ifx\csname bibnamefont\endcsname\relax
  \def\bibnamefont#1{#1}\fi
\expandafter\ifx\csname bibfnamefont\endcsname\relax
  \def\bibfnamefont#1{#1}\fi
\expandafter\ifx\csname citenamefont\endcsname\relax
  \def\citenamefont#1{#1}\fi
\expandafter\ifx\csname url\endcsname\relax
  \def\url#1{\texttt{#1}}\fi
\expandafter\ifx\csname urlprefix\endcsname\relax\def\urlprefix{URL }\fi
\providecommand{\bibinfo}[2]{#2}
\providecommand{\eprint}[2][]{\url{#2}}

\bibitem[{\citenamefont{Albert and Barab{\'a}si}(2002)}]{al02}
\bibinfo{author}{\bibfnamefont{R.}~\bibnamefont{Albert}} \bibnamefont{and}
  \bibinfo{author}{\bibfnamefont{A.~L.} \bibnamefont{Barab{\'a}si}},
  \bibinfo{journal}{Rev. Mod. Phys.} \textbf{\bibinfo{volume}{74}},
  \bibinfo{pages}{47} (\bibinfo{year}{2002}).

\bibitem[{\citenamefont{Newman}(2003)}]{ne03}
\bibinfo{author}{\bibfnamefont{M.~E.~J.} \bibnamefont{Newman}},
  \bibinfo{journal}{SIAM Rev.} \textbf{\bibinfo{volume}{45}},
  \bibinfo{pages}{167} (\bibinfo{year}{2003}).

\bibitem[{\citenamefont{Newman et~al.}(2006)\citenamefont{Newman, Barab\'asi,
  and Watts}}]{ne06}
\bibinfo{editor}{\bibfnamefont{M.~E.~J.} \bibnamefont{Newman}},
  \bibinfo{editor}{\bibfnamefont{A.~L.} \bibnamefont{Barab\'asi}},
  \bibnamefont{and} \bibinfo{editor}{\bibfnamefont{D.~J.} \bibnamefont{Watts}},
  eds., \emph{\bibinfo{title}{The structure and dynamics of networks}}
  (\bibinfo{publisher}{Princeton University}, \bibinfo{address}{Princeton},
  \bibinfo{year}{2006}).

\bibitem[{\citenamefont{Dorogovtsev and Mendes}(2003)}]{do03a}
\bibinfo{author}{\bibfnamefont{S.~N.} \bibnamefont{Dorogovtsev}}
  \bibnamefont{and} \bibinfo{author}{\bibfnamefont{J.~F.~F.}
  \bibnamefont{Mendes}}, \emph{\bibinfo{title}{Evolution of Networks: From
  Biological Nets to the Internet and WWW}} (\bibinfo{publisher}{Oxford
  University}, \bibinfo{address}{Oxford}, \bibinfo{year}{2003}).

\bibitem[{\citenamefont{da~F.~Costa et~al.}(2007)\citenamefont{da~F.~Costa,
  Rodrigues, Travieso, and {P. R. Villas Boas}}}]{co07}
\bibinfo{author}{\bibfnamefont{L.}~\bibnamefont{da~F.~Costa}},
  \bibinfo{author}{\bibfnamefont{F.~A.} \bibnamefont{Rodrigues}},
  \bibinfo{author}{\bibfnamefont{G.}~\bibnamefont{Travieso}}, \bibnamefont{and}
  \bibinfo{author}{\bibnamefont{{P. R. Villas Boas}}}, \bibinfo{journal}{Adv.
  Phys.} \textbf{\bibinfo{volume}{56}}, \bibinfo{pages}{167}
  (\bibinfo{year}{2007}).

\bibitem[{\citenamefont{Watts and Strogatz}(1998)}]{wa98}
\bibinfo{author}{\bibfnamefont{D.~J.} \bibnamefont{Watts}} \bibnamefont{and}
  \bibinfo{author}{\bibfnamefont{S.~H.} \bibnamefont{Strogatz}},
  \bibinfo{journal}{Nature} \textbf{\bibinfo{volume}{393}},
  \bibinfo{pages}{440} (\bibinfo{year}{1998}).

\bibitem[{\citenamefont{Barab{\'a}si and Albert}(1999)}]{ba99b}
\bibinfo{author}{\bibfnamefont{A.~L.} \bibnamefont{Barab{\'a}si}}
  \bibnamefont{and} \bibinfo{author}{\bibfnamefont{R.}~\bibnamefont{Albert}},
  \bibinfo{journal}{Science} \textbf{\bibinfo{volume}{286}},
  \bibinfo{pages}{509} (\bibinfo{year}{1999}).

\bibitem[{\citenamefont{Dorogovtsev and Mendes}(2002)}]{do02a}
\bibinfo{author}{\bibfnamefont{S.~N.} \bibnamefont{Dorogovtsev}}
  \bibnamefont{and} \bibinfo{author}{\bibfnamefont{J.~F.~F.}
  \bibnamefont{Mendes}}, \bibinfo{journal}{Adv. Phys.}
  \textbf{\bibinfo{volume}{51}}, \bibinfo{pages}{1079} (\bibinfo{year}{2002}).

\bibitem[{\citenamefont{Goh et~al.}(2002)\citenamefont{Goh, Oh, Jeong, Kahng,
  and Kim}}]{go02}
\bibinfo{author}{\bibfnamefont{K.~I.} \bibnamefont{Goh}},
  \bibinfo{author}{\bibfnamefont{E.~S.} \bibnamefont{Oh}},
  \bibinfo{author}{\bibfnamefont{H.}~\bibnamefont{Jeong}},
  \bibinfo{author}{\bibfnamefont{B.}~\bibnamefont{Kahng}}, \bibnamefont{and}
  \bibinfo{author}{\bibfnamefont{D.}~\bibnamefont{Kim}},
  \bibinfo{journal}{Proc. Natl. Acad. Sci. USA} \textbf{\bibinfo{volume}{99}},
  \bibinfo{pages}{12583} (\bibinfo{year}{2002}).

\bibitem[{\citenamefont{Siganos et~al.}(2003)\citenamefont{Siganos, Faloutsos,
  Faloutsos, and Faloutsos}}]{si03}
\bibinfo{author}{\bibfnamefont{G.}~\bibnamefont{Siganos}},
  \bibinfo{author}{\bibfnamefont{M.}~\bibnamefont{Faloutsos}},
  \bibinfo{author}{\bibfnamefont{P.}~\bibnamefont{Faloutsos}},
  \bibnamefont{and}
  \bibinfo{author}{\bibfnamefont{C.}~\bibnamefont{Faloutsos}},
  \bibinfo{journal}{IEEE ACM Trans. Netw.} \textbf{\bibinfo{volume}{11}},
  \bibinfo{pages}{514} (\bibinfo{year}{2003}).

\bibitem[{\citenamefont{Albert et~al.}(1999)\citenamefont{Albert, Jeong, and
  Barab{\'a}si}}]{al99}
\bibinfo{author}{\bibfnamefont{R.}~\bibnamefont{Albert}},
  \bibinfo{author}{\bibfnamefont{H.}~\bibnamefont{Jeong}}, \bibnamefont{and}
  \bibinfo{author}{\bibfnamefont{A.~L.} \bibnamefont{Barab{\'a}si}},
  \bibinfo{journal}{Nature} \textbf{\bibinfo{volume}{401}},
  \bibinfo{pages}{130} (\bibinfo{year}{1999}).

\bibitem[{\citenamefont{Jeong et~al.}(2001)\citenamefont{Jeong, Mason,
  Barab{\'a}si, and Oltvai}}]{je01}
\bibinfo{author}{\bibfnamefont{H.}~\bibnamefont{Jeong}},
  \bibinfo{author}{\bibfnamefont{S.~P.} \bibnamefont{Mason}},
  \bibinfo{author}{\bibfnamefont{A.~L.} \bibnamefont{Barab{\'a}si}},
  \bibnamefont{and} \bibinfo{author}{\bibfnamefont{Z.~N.}
  \bibnamefont{Oltvai}}, \bibinfo{journal}{Nature}
  \textbf{\bibinfo{volume}{411}}, \bibinfo{pages}{41} (\bibinfo{year}{2001}).

\bibitem[{\citenamefont{Newman}(2001)}]{ne01a}
\bibinfo{author}{\bibfnamefont{M.~E.~J.} \bibnamefont{Newman}},
  \bibinfo{journal}{Proc. Natl. Acad. Sci. USA} \textbf{\bibinfo{volume}{98}},
  \bibinfo{pages}{404} (\bibinfo{year}{2001}).

\bibitem[{\citenamefont{Bogacz et~al.}(2006)\citenamefont{Bogacz, Burda, and
  Waclaw}}]{bo06}
\bibinfo{author}{\bibfnamefont{L.}~\bibnamefont{Bogacz}},
  \bibinfo{author}{\bibfnamefont{Z.}~\bibnamefont{Burda}}, \bibnamefont{and}
  \bibinfo{author}{\bibfnamefont{B.}~\bibnamefont{Waclaw}},
  \bibinfo{journal}{Physica A} \textbf{\bibinfo{volume}{366}},
  \bibinfo{pages}{587} (\bibinfo{year}{2006}).

\bibitem[{\citenamefont{Barrat and Weigt}(2000)}]{ba00}
\bibinfo{author}{\bibfnamefont{A.}~\bibnamefont{Barrat}} \bibnamefont{and}
  \bibinfo{author}{\bibfnamefont{M.}~\bibnamefont{Weigt}},
  \bibinfo{journal}{Eur. Phys. J. B} \textbf{\bibinfo{volume}{13}},
  \bibinfo{pages}{547} (\bibinfo{year}{2000}).

\bibitem[{\citenamefont{Svenson and Johnston}(2002)}]{sv02}
\bibinfo{author}{\bibfnamefont{P.}~\bibnamefont{Svenson}} \bibnamefont{and}
  \bibinfo{author}{\bibfnamefont{D.~A.} \bibnamefont{Johnston}},
  \bibinfo{journal}{Phys. Rev. E} \textbf{\bibinfo{volume}{65}},
  \bibinfo{pages}{036105} (\bibinfo{year}{2002}).

\bibitem[{\citenamefont{Herrero}(2002)}]{he02a}
\bibinfo{author}{\bibfnamefont{C.~P.} \bibnamefont{Herrero}},
  \bibinfo{journal}{Phys. Rev. E} \textbf{\bibinfo{volume}{65}},
  \bibinfo{pages}{066110} (\bibinfo{year}{2002}).

\bibitem[{\citenamefont{{J. Viana Lopes} et~al.}(2004)\citenamefont{{J. Viana
  Lopes}, Pogorelov, {J. M. B. Lopes dos Santos}, and Toral}}]{lo04}
\bibinfo{author}{\bibnamefont{{J. Viana Lopes}}},
  \bibinfo{author}{\bibfnamefont{Y.~G.} \bibnamefont{Pogorelov}},
  \bibinfo{author}{\bibnamefont{{J. M. B. Lopes dos Santos}}},
  \bibnamefont{and} \bibinfo{author}{\bibfnamefont{R.}~\bibnamefont{Toral}},
  \bibinfo{journal}{Phys. Rev. E} \textbf{\bibinfo{volume}{70}},
  \bibinfo{pages}{026112} (\bibinfo{year}{2004}).

\bibitem[{\citenamefont{Candia}(2006)}]{ca06}
\bibinfo{author}{\bibfnamefont{J.}~\bibnamefont{Candia}},
  \bibinfo{journal}{Phys. Rev. E} \textbf{\bibinfo{volume}{74}},
  \bibinfo{pages}{031101} (\bibinfo{year}{2006}).

\bibitem[{\citenamefont{Dorogovtsev et~al.}(2008)\citenamefont{Dorogovtsev,
  Goltsev, and Mendes}}]{do08}
\bibinfo{author}{\bibfnamefont{S.~N.} \bibnamefont{Dorogovtsev}},
  \bibinfo{author}{\bibfnamefont{A.~V.} \bibnamefont{Goltsev}},
  \bibnamefont{and} \bibinfo{author}{\bibfnamefont{J.~F.~F.}
  \bibnamefont{Mendes}}, \bibinfo{journal}{Rev. Mod. Phys.}
  \textbf{\bibinfo{volume}{80}}, \bibinfo{pages}{1275} (\bibinfo{year}{2008}).

\bibitem[{\citenamefont{Igl\'oi and Turban}(2002)}]{ig02}
\bibinfo{author}{\bibfnamefont{F.}~\bibnamefont{Igl\'oi}} \bibnamefont{and}
  \bibinfo{author}{\bibfnamefont{L.}~\bibnamefont{Turban}},
  \bibinfo{journal}{Phys. Rev. E} \textbf{\bibinfo{volume}{66}},
  \bibinfo{pages}{036140} (\bibinfo{year}{2002}).

\bibitem[{\citenamefont{Dorogovtsev et~al.}(2002)\citenamefont{Dorogovtsev,
  Goltsev, and Mendes}}]{do02b}
\bibinfo{author}{\bibfnamefont{S.~N.} \bibnamefont{Dorogovtsev}},
  \bibinfo{author}{\bibfnamefont{A.~V.} \bibnamefont{Goltsev}},
  \bibnamefont{and} \bibinfo{author}{\bibfnamefont{J.~F.~F.}
  \bibnamefont{Mendes}}, \bibinfo{journal}{Phys. Rev. E}
  \textbf{\bibinfo{volume}{66}}, \bibinfo{pages}{016104}
  (\bibinfo{year}{2002}).

\bibitem[{\citenamefont{Leone et~al.}(2002)\citenamefont{Leone, V\'azquez,
  Vespignani, and Zecchina}}]{le02}
\bibinfo{author}{\bibfnamefont{M.}~\bibnamefont{Leone}},
  \bibinfo{author}{\bibfnamefont{A.}~\bibnamefont{V\'azquez}},
  \bibinfo{author}{\bibfnamefont{A.}~\bibnamefont{Vespignani}},
  \bibnamefont{and} \bibinfo{author}{\bibfnamefont{R.}~\bibnamefont{Zecchina}},
  \bibinfo{journal}{Eur. Phys. J. B} \textbf{\bibinfo{volume}{28}},
  \bibinfo{pages}{191} (\bibinfo{year}{2002}).

\bibitem[{\citenamefont{Herrero}(2004)}]{he04}
\bibinfo{author}{\bibfnamefont{C.~P.} \bibnamefont{Herrero}},
  \bibinfo{journal}{Phys. Rev. E} \textbf{\bibinfo{volume}{69}},
  \bibinfo{pages}{067109} (\bibinfo{year}{2004}).

\bibitem[{\citenamefont{Mydosh}(1993)}]{my93}
\bibinfo{author}{\bibfnamefont{J.~A.} \bibnamefont{Mydosh}},
  \emph{\bibinfo{title}{Spin Glasses. An Experimental Introduction}}
  (\bibinfo{publisher}{Taylor \& Francis}, \bibinfo{address}{London},
  \bibinfo{year}{1993}).

\bibitem[{\citenamefont{Fischer and Hertz}(1991)}]{fi91}
\bibinfo{author}{\bibfnamefont{K.~H.} \bibnamefont{Fischer}} \bibnamefont{and}
  \bibinfo{author}{\bibfnamefont{J.~A.} \bibnamefont{Hertz}},
  \emph{\bibinfo{title}{Spin Glasses}} (\bibinfo{publisher}{Cambridge
  University}, \bibinfo{address}{Cambridge}, \bibinfo{year}{1991}).

\bibitem[{\citenamefont{Kanter and Sompolinsky}(1987)}]{ka87}
\bibinfo{author}{\bibfnamefont{I.}~\bibnamefont{Kanter}} \bibnamefont{and}
  \bibinfo{author}{\bibfnamefont{H.}~\bibnamefont{Sompolinsky}},
  \bibinfo{journal}{Phys. Rev. Lett.} \textbf{\bibinfo{volume}{58}},
  \bibinfo{pages}{164} (\bibinfo{year}{1987}).

\bibitem[{\citenamefont{Dean and Lef\`evre}(2001)}]{de01}
\bibinfo{author}{\bibfnamefont{D.~S.} \bibnamefont{Dean}} \bibnamefont{and}
  \bibinfo{author}{\bibfnamefont{A.}~\bibnamefont{Lef\`evre}},
  \bibinfo{journal}{Phys. Rev. Lett.} \textbf{\bibinfo{volume}{86}},
  \bibinfo{pages}{5639} (\bibinfo{year}{2001}).

\bibitem[{\citenamefont{Kim et~al.}(2005)\citenamefont{Kim, Rodgers, Kahng, and
  Kim}}]{ki05}
\bibinfo{author}{\bibfnamefont{D.-H.} \bibnamefont{Kim}},
  \bibinfo{author}{\bibfnamefont{G.~J.} \bibnamefont{Rodgers}},
  \bibinfo{author}{\bibfnamefont{B.}~\bibnamefont{Kahng}}, \bibnamefont{and}
  \bibinfo{author}{\bibfnamefont{D.}~\bibnamefont{Kim}},
  \bibinfo{journal}{Phys. Rev. E} \textbf{\bibinfo{volume}{71}},
  \bibinfo{pages}{056115} (\bibinfo{year}{2005}).

\bibitem[{\citenamefont{Boettcher}(2003)}]{bo03}
\bibinfo{author}{\bibfnamefont{S.}~\bibnamefont{Boettcher}},
  \bibinfo{journal}{Phys. Rev. B} \textbf{\bibinfo{volume}{67}},
  \bibinfo{pages}{060403(R)} (\bibinfo{year}{2003}).

\bibitem[{\citenamefont{Nikoletopoulos
  et~al.}(2004)\citenamefont{Nikoletopoulos, Coolen, Castillo, Skantzos,
  Hatchett, and Wemmenhove}}]{ni04}
\bibinfo{author}{\bibfnamefont{T.}~\bibnamefont{Nikoletopoulos}},
  \bibinfo{author}{\bibfnamefont{A.~C.~C.} \bibnamefont{Coolen}},
  \bibinfo{author}{\bibfnamefont{I.~P.} \bibnamefont{Castillo}},
  \bibinfo{author}{\bibfnamefont{N.~S.} \bibnamefont{Skantzos}},
  \bibinfo{author}{\bibfnamefont{J.~P.~L.} \bibnamefont{Hatchett}},
  \bibnamefont{and}
  \bibinfo{author}{\bibfnamefont{B.}~\bibnamefont{Wemmenhove}},
  \bibinfo{journal}{J. Phys. A} \textbf{\bibinfo{volume}{37}},
  \bibinfo{pages}{6455} (\bibinfo{year}{2004}).

\bibitem[{\citenamefont{Wemmenhove et~al.}(2005)\citenamefont{Wemmenhove,
  Nikoletopoulos, and Hatchett}}]{we05}
\bibinfo{author}{\bibfnamefont{B.}~\bibnamefont{Wemmenhove}},
  \bibinfo{author}{\bibfnamefont{T.}~\bibnamefont{Nikoletopoulos}},
  \bibnamefont{and} \bibinfo{author}{\bibfnamefont{J.~P.~L.}
  \bibnamefont{Hatchett}}, \bibinfo{journal}{J. Stat. Mech.: Theory Exp.} p.
  \bibinfo{pages}{P11007} (\bibinfo{year}{2005}).

\bibitem[{\citenamefont{Weigel and Johnston}(2007)}]{we07}
\bibinfo{author}{\bibfnamefont{M.}~\bibnamefont{Weigel}} \bibnamefont{and}
  \bibinfo{author}{\bibfnamefont{D.}~\bibnamefont{Johnston}},
  \bibinfo{journal}{Phys. Rev. B} \textbf{\bibinfo{volume}{76}},
  \bibinfo{pages}{054408} (\bibinfo{year}{2007}).

\bibitem[{\citenamefont{Ostilli and Mendes}(2008)}]{os08}
\bibinfo{author}{\bibfnamefont{M.}~\bibnamefont{Ostilli}} \bibnamefont{and}
  \bibinfo{author}{\bibfnamefont{J.~F.~F.} \bibnamefont{Mendes}},
  \bibinfo{journal}{Phys. Rev. E} \textbf{\bibinfo{volume}{78}},
  \bibinfo{pages}{031102} (\bibinfo{year}{2008}).

\bibitem[{\citenamefont{Bartolozzi et~al.}(2006)\citenamefont{Bartolozzi,
  Surungan, Leinweber, and Williams}}]{ba06}
\bibinfo{author}{\bibfnamefont{M.}~\bibnamefont{Bartolozzi}},
  \bibinfo{author}{\bibfnamefont{T.}~\bibnamefont{Surungan}},
  \bibinfo{author}{\bibfnamefont{D.~B.} \bibnamefont{Leinweber}},
  \bibnamefont{and} \bibinfo{author}{\bibfnamefont{A.~G.}
  \bibnamefont{Williams}}, \bibinfo{journal}{Phys. Rev. B}
  \textbf{\bibinfo{volume}{73}}, \bibinfo{pages}{224419}
  (\bibinfo{year}{2006}).

\bibitem[{\citenamefont{Herrero}(2008)}]{he08}
\bibinfo{author}{\bibfnamefont{C.~P.} \bibnamefont{Herrero}},
  \bibinfo{journal}{Phys. Rev. E} \textbf{\bibinfo{volume}{77}},
  \bibinfo{pages}{041102} (\bibinfo{year}{2008}).

\bibitem[{\citenamefont{Newman}(2005)}]{ne05}
\bibinfo{author}{\bibfnamefont{M.~E.~J.} \bibnamefont{Newman}},
  \bibinfo{journal}{Comtemp. Phys.} \textbf{\bibinfo{volume}{46}},
  \bibinfo{pages}{323} (\bibinfo{year}{2005}).

\bibitem[{\citenamefont{Krawczyk et~al.}(2005)\citenamefont{Krawczyk, Malarz,
  Kawecka-Magiera, Maksymowicz, and Kulakowski}}]{kr05}
\bibinfo{author}{\bibfnamefont{M.~J.} \bibnamefont{Krawczyk}},
  \bibinfo{author}{\bibfnamefont{K.}~\bibnamefont{Malarz}},
  \bibinfo{author}{\bibfnamefont{B.}~\bibnamefont{Kawecka-Magiera}},
  \bibinfo{author}{\bibfnamefont{A.~Z.} \bibnamefont{Maksymowicz}},
  \bibnamefont{and}
  \bibinfo{author}{\bibfnamefont{K.}~\bibnamefont{Kulakowski}},
  \bibinfo{journal}{Phys. Rev. B} \textbf{\bibinfo{volume}{72}},
  \bibinfo{pages}{024445} (\bibinfo{year}{2005}).

\bibitem[{\citenamefont{Binder and Heermann}(1997)}]{bi97}
\bibinfo{author}{\bibfnamefont{K.}~\bibnamefont{Binder}} \bibnamefont{and}
  \bibinfo{author}{\bibfnamefont{D.}~\bibnamefont{Heermann}},
  \emph{\bibinfo{title}{Monte Carlo Simulation in Statistical Physics}}
  (\bibinfo{publisher}{Springer-Verlag}, \bibinfo{address}{Berlin},
  \bibinfo{year}{1997}).

\bibitem[{\citenamefont{Bollob\'as}(1998)}]{bo98}
\bibinfo{author}{\bibfnamefont{B.}~\bibnamefont{Bollob\'as}},
  \emph{\bibinfo{title}{Modern Graph Theory}}
  (\bibinfo{publisher}{Springer-Verlag}, \bibinfo{address}{New York},
  \bibinfo{year}{1998}).

\bibitem[{\citenamefont{Parisi}(1983)}]{pa83}
\bibinfo{author}{\bibfnamefont{G.}~\bibnamefont{Parisi}},
  \bibinfo{journal}{Phys. Rev. Lett.} \textbf{\bibinfo{volume}{50}},
  \bibinfo{pages}{1946} (\bibinfo{year}{1983}).

\bibitem[{\citenamefont{Kawashima and Young}(1996)}]{ka96}
\bibinfo{author}{\bibfnamefont{N.}~\bibnamefont{Kawashima}} \bibnamefont{and}
  \bibinfo{author}{\bibfnamefont{A.~P.} \bibnamefont{Young}},
  \bibinfo{journal}{Phys. Rev. B} \textbf{\bibinfo{volume}{53}},
  \bibinfo{pages}{R484} (\bibinfo{year}{1996}).

\bibitem[{\citenamefont{Katzgraber and Young}(2002)}]{ka02}
\bibinfo{author}{\bibfnamefont{H.~G.} \bibnamefont{Katzgraber}}
  \bibnamefont{and} \bibinfo{author}{\bibfnamefont{A.~P.} \bibnamefont{Young}},
  \bibinfo{journal}{Phys. Rev. B} \textbf{\bibinfo{volume}{65}},
  \bibinfo{pages}{214402} (\bibinfo{year}{2002}).

\end{thebibliography}

\end{document}